\documentclass[aps,floatfix,showpacs,twocolumn]{revtex4}
\usepackage{natbib}
\usepackage{dcolumn}
\usepackage{graphicx}

\newcommand{\physrep}{Phys.~Rep.}

\newcommand{\be}{\begin{equation}}
\newcommand{\ee}{\end{equation}}
\newcommand{\bea}{\begin{eqnarray}}
\newcommand{\eea}{\end{eqnarray}}
\begin{document}
\title{Light Deflection, Lensing, and Time Delays from Gravitational Potentials and Fermat's Principle in the Presence of a Cosmological Constant}
\author{Mustapha Ishak\footnote{Electronic address: mishak@utdallas.edu}}
\affiliation{
Department of Physics, The University of Texas at Dallas, Richardson, TX 75083, USA}
\date{\today}
\begin{abstract}
The contribution of the cosmological constant to the deflection angle and the time delays are derived from the integration of the gravitational potential as well as from Fermat's Principle. The findings are in agreement with recent results using exact solutions to Einstein's equations and reproduce precisely the new $\Lambda$-term in the bending angle and the lens equation. The consequences on time delay expressions are explored. While it is known that $\Lambda$ contributes to the gravitational time delay, it is shown here that a new $\Lambda$-term appears in the geometrical time delay as well. Although these newly derived terms are perhaps small for current observations, they do not cancel out as previously claimed. Moreover, as shown before, at galaxy cluster scale, the $\Lambda$ contribution can be larger than the second-order term in the Einstein deflection angle for several cluster lens systems.
\end{abstract}
\maketitle
\section{introduction}
Various complementary observations seem to indicate that the expansion of the universe has entered a phase of acceleration \cite{observations}. This has become one of the most important current problems in cosmology and all physics, as discussed, for example, in {\cite{reviews} and references therein. While several explanations are possible \cite{reviews}, current observations \cite{observations} are consistent with a cosmological constant in the Einstein Field Equations. 

In a recent work, the authors of \cite{RindlerAndIshak2007} used the Schwarzschild-de Sitter (SdS) metric and  showed that the cosmological constant does contribute to the bending of light around a concentrated mass. It was shown there that even though $\Lambda$ drops out from the null geodesic equation, it still contributes to the deflection of light \textit{simply because of the geometry of the SdS spacetime} represented by the metric, regardless of other considerations. The result made a correction to the question (see e.g. \cite{Islam,Freire,Kagramanova,Finelli,Sereno}) and was confirmed in \cite{Lake2007,Sereno2007,Schucker}. Next, in reference \cite{Ishaketal2007}, the authors brought the result into an observational context and used an exact solution construction where a Schwarzschild-de Sitter vacuole was embedded into a Friedmann-Lemaitre-Robertson-Walker (FLRW) background, and where a $\Lambda$-term was added to the deflection angle within the broadly used lens equation. Finally, using observations of Einstein angles around clusters, the authors were able to put  upper-bounds on the value of the cosmological constant, only two orders of magnitude away from the value determined by cosmological probe constraints.

In this paper, we use the integration of the gravitational potential method and also Fermat's Principle to calculate the contribution of the cosmological constant to the deflection of light, the lens equation, and the time delays. Whereas it is known that $\Lambda$ contributes to the gravitational time delay, we show here that a new $\Lambda$-term appears in the geometrical time delay.

First, It is worth clarifying a point that has been a source of ambiguity in some of the recent literature about the effect of $\Lambda$ on the bending of light and gravitational lensing by a galaxy cluster in an FLRW spacetime. As we show in Figure 1, the light bending due the lens occurs in a region close to the lens (SdS vacuole) and then once the light transitions out of the vacuole into FLRW spacetime then the bending stops. Away from the cluster (i.e. outside the vacuole), the ray of light propagates in an FLRW background. Now, from the geometry, the lens equation includes the bending angle and also angular diameter distances. While $\Lambda$ is present in the lens equation via the usual expressions for the angular diameter distances in FLRW, $\Lambda$ also enters the lens equation via the bending angle itself that occurs inside the SdS vacuole. As evaluated in \cite{Ishaketal2007}, the contribution of $\Lambda$ to the deflection angle is small but larger than the second-order term in the Einstein angle for several cluster lens systems. Moreover, this is also a matter of mathematical correctness since the newly derived terms do not cancel out as previously claimed. In relation to this clarification about the lens geometry above, recent papers \cite{Khriplovich,Sereno2007} refer to our previous work \cite{RindlerAndIshak2007,Ishaketal2007} without taking into consideration the construction described above and used in \cite{Ishaketal2007}. Then, it is concluded in \cite{Khriplovich} that terms of the order of $\sim\Lambda r_g r_0$ (in their notation) may exist. That term is precisely the $\Lambda$-term derived in our previous work \cite{Ishaketal2007}. It is hopped that this paragraph along with Figure 1 will clarify this point about the cosmological constant and the lens equation geometry.

\begin{figure*}
\begin{center}
\includegraphics[width=6.6in,height=2.2in,angle=0]{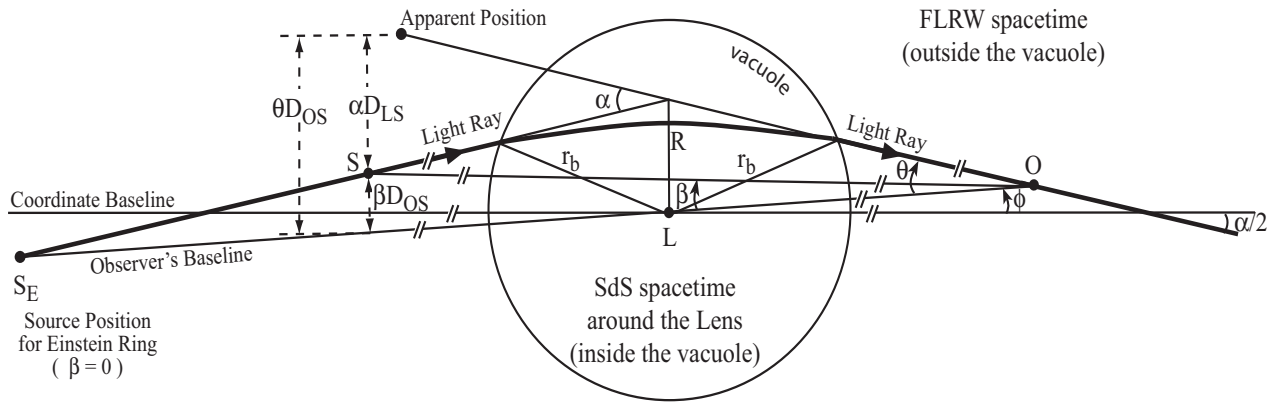}
\caption{\label{fig:figure1} 
The lens equation geometry. Observer, lens, and source are at O, L, and S, respectively. The position of the unlensed source is at an angle $\beta$, the apparent position is at the angle $\theta$ and the deflection angle is $\alpha$. The distance from the observer to the source is $D_{OS}$, from the observer to the lens is $D_{OL}$, and from the lens to the source is $D_{LS}$. As usual, the lens equation follows from the geometry as $\theta D_{OS}=\beta D_{OS} + \alpha D_{LS}$. A ray of light propagates from the left within the FLRW background, then enters inside the Schwarzschild-de Sitter vacuole around the lens where it gets deflected by the lens, and then it exits the vacuole and continues its propagation in the FLRW background to the right. The results derived from matching exact solutions to Einstein's equations in \cite{Ishaketal2007} and those derived in this work from the usual approximate constructions used in gravitational lensing literature are in perfect agreement for the contribution of $\Lambda$ to the deflection angle and the lens equation. The same result is also derived using Fermat's Principle.  
}
\end{center}
\end{figure*}

It was recently pointed out in \cite{Sereno2008b} that some of the $\Lambda$-terms contributing to the deflection angle could be incorporated into the angular diameter distances when these are extended up to the lens and not the boundary of the vacuole while other $\Lambda$-terms cannot; this will be investigated elsewhere.   

It is also worth noting that the contribution of the cosmological constant to the bending angle enters via the Weyl focusing of the Sachs equations \cite{Sachs1961,Dyer1977,Schneideretal1992} and takes place inside the SdS vacuole (see also footnote [22] in previous paper \cite{RindlerAndIshak2007}), while outside the vacuole only the Ricci focusing \cite{Sachs1961,Dyer1977,Schneideretal1992} takes place and the cosmological constant appears via the angular distances in the FLRW background. This applies to the bending angle and the related time-delay as well. 

As we will show, the $\Lambda$-term in the lens equation as discussed in \cite{Ishaketal2007} can be derived using various methods used in gravitational lensing and propagates to time delay calculations.

\section{Deflection potentials, the lens equation, and the cosmological constant}

In this section, we show that the contribution of the cosmological constant to the light bending angle \cite{RindlerAndIshak2007} and the lens equation \cite{Ishaketal2007} can also be derived from gravitational potentials, a method that is frequently used in gravitational lensing literature, see for example \cite{Schneideretal1992,Bartelmann,Pyne,Carroll}. 

Ishak, Rindler \textit{et al.} derived in \cite{Ishaketal2007} an amended lens equation where the contribution of the cosmological constant was added to the Einstein bending angle. The $\Lambda$-term was derived there using an exact construction of a Schwarzschild-de Sitter (SdS) vacuole embedded in a Friedmann-Lemaitre-Robertson-Walker (FLRW) background and was given by  
\be
\alpha_{\Lambda}=-\frac{\Lambda\,R\,r_b}{3},
\label{eq:Ishak-Rindler-Term}
\ee
where $R$ is a radius close to the lens as used in \cite{RindlerAndIshak2007,Ishaketal2007} and similar to the impact parameter, e.g. \cite{MTW,Wald1984}, but distinct from it since the SdS spacetime is not asymptotically flat, and $r_b$ is the radial coordinate at the boundary of the vacuole where the spacetime transitions from the SdS spacetime to an FLRW background. 

We note that one can use again the following matching conditions at the boundary of the vacuole, see e.g.  \cite{Darmois,Israel,Swiss-cheese},  
\be
r_{b\,\, in \, SdS} = a(t) \,\,\textit{r}_{b\,\,in\,FLRW}
\label{eq:cond1}
\ee
and
\be
m_{\,SdS} = \frac{4 \pi}{3}\,\, r_{b\,\, in \, SdS}^3\times \rho_{matter\,in\,FLRW},        
\label{eq:cond2}
\ee
and re-write the Ishak-Rindler $\Lambda$-term, (\ref{eq:Ishak-Rindler-Term}), as 
\be
\alpha_{\Lambda}= -\frac{\Lambda\,R}{3}\left( \frac{3\,m_{\,SdS}}{4 \pi \,\rho_{matter\,in\,FLRW} } \right) ^{(1/3)}.
\ee
This expression gives the $\Lambda$-term as a function of the mass of the galaxy or cluster lens in the SdS vacuole and the matter density of the FLRW background.

The second method that we shall employ to calculate $\alpha_{\Lambda}$ is analogous to the method of approximation that is frequently used in gravitational lensing literature where the lens (inhomogeneity) in an FLRW background is represented by a Newtonian potential inserted  in a post-Minkowskian line element or a post-FLRW line element (see for example \cite{Schneideretal1992,Bartelmann,Pyne,Carroll}). The metric in such a construction is then given by $g_{ab}=\eta_{ab}+h_{ab}$ where $h_{ab}$ measure the departure from the Minkowskian metric $\eta_{ab}$.

First, we recall that $h_{ab}$ in linearized Schwarzschild spacetime and the associated Newtonian potential $\Phi$ can be read off the usual Schwarzschild metric in isotropic coordinates given by (e.g. \cite{Rindler,MTW}) (we use relativistic units where we set $G=c=1$)
\be
ds^2=-\left(\frac{1-\frac{m}{2r}}{1+ \frac{m}{2r}}\right)^2 dt^2 + \left(1+\frac{m}{2r}\right)^4(dr^2+r^2 d\Omega^2),
\ee
and are given by 
\be
\Phi=-\frac{h_{tt}}{2}=-\frac{h_{ii}}{2}=-\frac{m}{r}.
\label{eq:Schw_Pert}
\ee
For a ray of light traveling in the $x$-direction, the first-order term of the Einstein deflection angle is then given by
\bea
\alpha&=&-\frac{1}{2}\,\int^{+x_b}_{-x_b} \nabla_{\perp} (h_{tt}+h_{xx}) dx \nonumber \\
      &=&  2\, \int^{+x_b}_{-x_b} \nabla_{\perp} \Phi(x,y,z) dx
\label{eq:secondmethod}
\eea
where $\nabla_{\perp}\equiv\nabla-\nabla_{\parallel}$ is the gradient transverse to the path, see for example \cite{Schneideretal1992,Carroll}. At the vacuole boundary, $x_b=\sqrt{r_b^2-R^2}$ and a straightforward integration yields, the Einstein first-order term  
\be
\alpha_{_{Einstein}}=\frac{4m}{R}. 
\label{eq:Einstein}
\ee
For the contribution of $\Lambda$ to the bending angle which happens close to the lens (i.e. inside the vacuole region), some subtleties come into consideration as we shall see. 
In linearized GR, we can add the $h_{ab}$ due to various sources. It therefore follows that the bending angle itself can be obtained by adding the bending angles due to different $h_{ab}$. Moreover, in order to comply with the isotropic FLRW background outside the vacuole, we should here also use isotropic coordinates for the de Sitter metric inside the vacuole, see e.g. \cite{Kerr,Klioner} and reference therein,
\be
ds^2=-\left(\frac{1-\frac{\Lambda r^2}{12}}{1+ \frac{\Lambda r^2}{12}}\right)^2 dt^2 + \frac{1}{(1+\frac{\Lambda r^2}{12})^2}(dr^2+r^2 d\Omega^2)
\ee
which linearizes to $g_{ab}=\eta_{ab}+h_{ab}$ with, see for example \cite{Kerr},
\bea
h_{tt}& = &-2\Phi = \frac{\Lambda r^2}{3} \nonumber \\ 
h_{0i}& = &0  \nonumber \\
h_{ij}& = &-2\Psi \delta_{ij}= -\frac{\Lambda r^2}{6}\delta_{ij}. 
\label{eq:h_Lambda}
\eea

The $\Lambda$ contribution to the bending angle is then found by using again 
\bea
\alpha_{\Lambda}&=&-\frac{1}{2}\,\int^{+x_b}_{-x_b} \nabla_{\perp} (h_{tt}+h_{xx}) dx \nonumber \\
      &=&  \, \int^{+x_b}_{-x_b} \nabla_{\perp} (\Phi+\Psi) dx = -\frac{\Lambda\,R\,r_b}{3}.
\label{eq:secondmethod}
\eea

Now, combining equations (\ref{eq:Einstein}) and (\ref{eq:secondmethod}), using the lens geometry (Figure 1), and the usual assumption of small angles, the lens equation reads 
\be
\theta-\beta=(\alpha_{Einstein}+\alpha_{\Lambda}) \,\frac{D_{LS}}{D_{OS}}.
\label{eq:lensequation}
\ee

The $\Lambda$-term (\ref{eq:secondmethod}) and the lens equation (\ref{eq:lensequation}) are exactly the $\Lambda$-term and the lens equation as derived in \cite{Ishaketal2007} from a different method based on the \textit{exact} matching of an SdS spacetime vacuole to an FLRW spacetime background \cite{Ishaketal2007}. 

In section (IV) below we shall give yet another alternative for deriving the $\Lambda$-term, this time from Fermat's Principle.

\section{Time Delays and the cosmological constant}

A ray of light traveling in the gravitational field is known to experience two types of time delay, see for example \cite{Schneideretal1992,Schechter}. The first is the geometrical time delay, $\tau_{geom}$, due to the extra path length resulting from the deflection. The second is the gravitational time delay, $\tau_{grav}$, due to the gravitational potential, known as the Shapiro delay \cite{Shapiro,Schneideretal1992}. The contribution of the cosmological constant to the gravitational time delay has been discussed in for example \cite{Kerr,Kagramanova}. However, we show here that $\Lambda$ also affects the geometrical time delay because of its contribution to the bending angle shortens the bent path. 

Before we derive the two terms, we want to clarify that the contributions of $\Lambda$ that we are interested in here are distinct from the obvious ones that come via the angular diameter distances as we explained in the introduction, but we are interested in the contributions that happen inside the vacuole around the lens.  

First, the gravitational time delay due to $\Lambda$ can be derived using (\ref{eq:h_Lambda}), see for example \cite{Kerr}, and is given by 
\bea
\tau_{\Lambda,grav}&=&\frac{1}{2}\,\int^{+x_b}_{-x_b} (h_{tt}+h_{xx}) dx \nonumber \\
      &=& \frac{\Lambda\,r_b^3}{18}+ \frac{\Lambda\,r_b\,R^2}{12}.  
\label{eq:TimeDelayLambda}
\eea
This result gives explicitly the two leading $\Lambda$-terms for the contribution of $\Lambda$ to the gravitational time delay. 
One can see that these terms are in agreement with \cite{Kerr} when their integration boundaries ${\mathbf{\rho_1}}$ and ${\mathbf{\rho_2}}$ are set equal to $+{\mathbf{r_b}}=(+x_b,R,0)$ and $-{\mathbf{r_b}}=(-x_b,R,0)$, respectively. 

Now, we recall the Shapiro time delay due to the Schwarzschild potential (\ref{eq:Schw_Pert}). It is given for our vacuole by 
\bea
\tau_{Shapiro}&=&  - 2\, \int^{+x_b}_{-x_b} \Phi_{m} dx = 4 m \log \left(\frac{2\,r_b}{R}\right).  
\label{eq:Shapiro}
\eea

So the gravitational time delay due to $\Lambda$ adds to the Shapiro delay. These time delays are due to the gravitational potentials which cause clocks to slow down \cite{Rindler,MTW}. 

For distant galaxies and clusters of galaxies, the Shapiro delay and the $\Lambda$-term need to take into account the redshift of the lens, the geometry as given by Figure 1, and the angular diameter distances as given in the FLRW metric away from the potential. Also, when dealing with measurements and observations, the gravitational time delays require modeling of the lens systems, see for example \cite{Schechter,Narayan1999}. Following the standard approach \cite{Schneideretal1992,Schechter}, we put the two time delays in a cosmological setting as
\be
\tau_{grav}=(1+z_{_L})\,\frac{D_{_{OS}}\,D_{_{OL}}}{D_{_{LS}}}\left(\tau_{Shapiro}+\tau_{\Lambda,grav}\right)
\ee
where all the quantities are as defined in Figure 1 and the angular-diameter distance is given by 
\be
D(z)=\frac{c}{H_0(1+z)}\int^{z}_{0}\frac{dz'}{\sqrt{\Omega_m(1+z')^3+\Omega_\Lambda}}
\ee
in a spatially flat cosmology. 

Next, we derive the geometrical time delay, $\tau_{\Lambda,geom}$, due to $\Lambda$. This term is new and results from the contribution of $\Lambda$ to the deflection angle and thus to the path length. A general derivation of $\tau_{geom}$ from the lens geometry (e.g. figure 1) and the FLRW metric can be found in for example \cite{Schneideretal1992}. The geometrical time delay is given by   
\be
\tau_{geom}=(1+z_{_L})\,\frac{D_{_{OS}}\,D_{_{OL}}}{D_{_{LS}}}\,\frac{1}{2}\,(\theta - \beta )^2.
\label{eq:geom_general}
\ee
We can use the lens equation (\ref{eq:lensequation}) to rewrite (\ref{eq:geom_general}) as 
\be
\tau_{geom}=(1+z_{_L})\,\frac{D_{_{LS}}\,D_{_{OL}}}{D_{_{OS}}}\,\frac{1}{2}\,\alpha^2.
\label{eq:geom_general2}
\ee
Now, we can expand $\alpha^2$ 
to write
\be
\tau_{geom}\approx (1+z_{_L})\,\frac{D_{_{LS}}\,D_{_{OL}}}{D_{_{OS}}}\,\frac{1}{2}\,(\alpha_{Einstein}^2+2\,\alpha_{Einstein}\,\alpha_{\Lambda})
\label{eq:geom_general3}
\ee
where we define the last term as
\bea
\tau_{\Lambda,geom}&\equiv& (1+z_{_L})\,\frac{D_{_{LS}}\,D_{_{OL}}}{D_{_{OS}}}\,\alpha_{Einstein}\,\,\alpha_{\Lambda}\nonumber \\ 
\nonumber \\
                   &=&(1+z_{_L})\,\frac{D_{_{LS}}\,D_{_{OL}}}{D_{_{OS}}}\,\left(-\frac{4\Lambda\,m\,r_b}{3}\right).
\label{eq:Lambda_geom}
\eea
The term (\ref{eq:Lambda_geom}) is new and represents the contribution of the cosmological constant to the geometrical time delay that enters from its contribution to the deflection angle. We clarify again that the contribution of $\Lambda$ as it appears in the last fraction in (\ref{eq:Lambda_geom}) is separate from the contributions that $\Lambda$ has via the angular diameter distances as given in the FLRW background, outside the vacuole. This newly derived term is perhaps observationally small but does not cancel out to zero as previously claimed.   

The sign of $\tau_{\Lambda,geom}$ is negative. So while the effect of a Schwarzschild mass is to produce a retardation due to the deflection of the light that it creates, the effect of $\Lambda$ on the geometrical time delay is the opposite. A positive cosmological constant diminishes the deflection angle and the length of the path that light needs to travel and hence it diminishes the geometrical time delay. 

Finally, we note that when using observations, it is rather the relative time delays between different images that are used \cite{Schechter,Narayan1999}, i.e.
\be
\Delta \tau_{\,1,2}=\tau_{\,image_1}-\tau_{\,image_2}.
\ee
Also, while observations fairly directly yield measurements of redshifts of lenses and deflection angles, time delay measurements require some sophisticated modeling of lens systems \cite{Schechter,Narayan1999}. These are out of the scope of this theoretical work but need to be studied in future work using modeling and simulations of observed cluster lens systems and images.  

We will show in the next section that the contribution of $\Lambda$ to the deflection angle and consequently to the geometrical time delay can also be derived from Fermat's Principle while the Shapiro delay follows directly from the spacetime metric.

\section{Fermat's Principle and the contribution of $\Lambda$ to the lens equation and the time delays}

The expression for the deflection angle also follows from using Fermat's Principle and the Euler-Lagrange equations of the variational Principle 
\be
\delta\int n\, dl 
\label{eq:Variation_of_N}
\ee
where $n$ is considered as an effective index of refraction of the gravitational field and $dl$ is the path of the ray of light.

First, as discussed in for example \cite{Schneideretal1992,Narayan1999}, the null curve $ds^2=g_{ab}dx^a dx^b=0$ for the Schwarzschild potential (\ref{eq:Schw_Pert}), leads to the Fermat's Principle (\ref{eq:Variation_of_N}), where
\be
n= 1- 2\Phi,
\label{eq;n_Schw}
\ee
from which the Euler-Lagrange equations give \cite{Schneideretal1992,Narayan1999},
\be
\alpha_{_{Einstein}}=2 \int \nabla_{\perp} \Phi dl=\frac{4m}{R}.
\ee

Now, we apply Fermat's Principle in order to derive the contribution of the cosmological constant to the bending angle and use $h_{ab}$ as given by (\ref{eq:h_Lambda}). For a \textit{future-directed} null curve,
\bea
ds^2&=&0 \nonumber \\
    &=&-dt^2\,(1-2\Phi)+(1-2\Psi)\,dl^2, \nonumber \\
    &= & -dt^2\,(1-\frac{\Lambda r^2}{3})+(1-\frac{\Lambda r^2}{6})\,dl^2,
\eea
or simply
\bea
dt&=&(1-(\Phi+\Psi))\,dl
  =(1+\frac{\Lambda r^2}{12})\,dl.
\label{eq:dte}
\eea
First, the gravitational time delay follows from the second term of the RHS of equation (\ref{eq:dte}). The integral of which gives 
\be
\Delta t = dl - \int (\Phi+\Psi) dl= dl + \tau_{\Lambda,grav}
\ee
where $\tau_{\Lambda,grav}$ was already calculated in (\ref{eq:TimeDelayLambda}) of the previous section.

Next, it follows from equations (\ref{eq:Variation_of_N}) and (\ref{eq:dte}) that 
\be
n=1-(\Phi+\Psi)=1+\frac{\Lambda r^2}{12}.
\label{eq:n_de_Sitter}
\ee

Following \cite{Schneideretal1992}, we consider a ray of light traveling along the path $dl$ with unit tangent vector $\mathbf{e}$. The deflection angle is given by the change in the direction of the null ray. From our equations (\ref{eq:Variation_of_N}) and (\ref{eq:n_de_Sitter}) and the Euler-Lagrange equations, it follows that
\bea
\frac{d\mathbf{e}}{dl}=-\left( \nabla (\Phi+\Psi)- \mathbf{e}\,(\mathbf{e}\,.\,\nabla (\Phi+\Psi)\right)=-\nabla_{\perp} (\Phi+\Psi) \nonumber \\
\eea
and
\bea
\mathbf{\alpha}_{\Lambda}=\mathbf{e}_{in}-\mathbf{e}_{out}=\int \nabla_{\perp} (\Phi+\Psi) dl,
\eea
a result already derived in (\ref{eq:secondmethod}) using null geodesics and the metric. So for a null ray traveling in the x-direction with integration boundaries as given in (\ref{eq:secondmethod}), this approach gives again the $\Lambda$-term as given in \cite{Ishaketal2007} and re-derived in section II, i.e. 
\be
\alpha_{\Lambda}=-\frac{\Lambda\,R\,r_b}{3}.
\ee

\section{Conclusion}
Using the integration of gravitational potentials and Fermat's Principle, we find that a cosmological constant term appears in the deflection angle and propagates to time delay calculations. We used here an approximate construction that is frequently used in gravitational lensing literature \cite{Schneideretal1992,Bartelmann} and where the lens (inhomogeneity) in an FLRW is represented by a potential inserted in a post-Minkowskian or post-FLRW line elements. Our results for the deflection angle and the lens equation are in perfect agreement with previous work \cite{Ishaketal2007,RindlerAndIshak2007} where an SdS spacetime vacuole was \textit{exactly embedded} in an FLRW spacetime background. Then we derived expressions for the effect of $\Lambda$ on the gravitational and geometrical time delays. Whereas it is known that the cosmological constant contributes to the gravitational time delay, we showed here that a \textit{new} $\Lambda$-term appears in the geometrical time delay. 

We also clarified the geometrical figure of interest whether one uses a construction based on exact solutions to Einstein's equations or based on approximate constructions. As shown, in Figure 1, the bending of light occurs in the region close to the deflector inside the SdS vacuole. Outside the vacuole, the ray of light simply propagates in an FLRW spacetime. The contribution of the cosmological constant to the bending angle enters via the Weyl focusing of the Sachs equations \cite{Sachs1961,Dyer1977,Schneideretal1992} and takes place inside the vacuole, while outside the vacuole only the Ricci focusing \cite{Sachs1961,Dyer1977,Schneideretal1992} takes place and $\Lambda$ appears via the angular distances in the FLRW background. 

Indeed, the $\Lambda$-terms that are derived here and in \cite{Ishaketal2007} are related to the bending of light which occurs inside the SdS vacuole, close to the lens.  These contributions are separate from the usual presence of $\Lambda$ in the expressions for angular diameter distances in the FLRW background.   It was recently discussed in reference \cite{Sereno2008b} that some of the $\Lambda$-terms contributing to the light bending angle could be incorporated into the angular diameter distances when these are extended up to the lens and not the boundary of the vacuole while other $\Lambda$-terms cannot, this will be investigated elsewhere. Although the newly derived $\Lambda$-terms for the lens equation and the time delay are perhaps small for current observations, \textit{they do not cancel out as previously claimed}. Moreover, at galaxy cluster scale, it was shown in \cite{Ishaketal2007} that the contribution of $\Lambda$ to the bending angle can be larger than the second-order term in the Einstein deflection angle for several cluster lens systems.

\acknowledgements 
The author thanks Wolfgang Rindler for useful discussions and Chris Allison for graphics work. The author acknowledges partial support from the Hoblitzelle Foundation.

\end{document}